\documentclass[10pt,aps,prb,reprint,amsfonts,amssymb,amsmath,floatfix,
longbibliography]{revtex4-2}
\usepackage{microtype}
\usepackage{lmodern}
\usepackage[utf8]{inputenc}
\usepackage[T1]{fontenc}
\usepackage{graphicx}
\usepackage{hyperref}
\hypersetup{
    colorlinks=true,
    linkcolor=blue,
    citecolor=blue,   
    urlcolor=blue
}
\begin{document}

\title{Momentum microscopy of Pb-intercalated graphene on SiC: charge neutrality and electronic structure of interfacial Pb}

\author{Bharti Matta}
\email{b.matta@fkf.mpg.de}
\author{Philipp Rosenzweig}
\email{p.rosenzweig@fkf.mpg.de}
\affiliation{Max-Planck-Institut für Festkörperforschung, Heisenbergstraße 1, 70569 Stuttgart, Germany}
\author{Olaf Bolkenbaas}
\altaffiliation[Present address: ]{Department of Applied Physics, Eindhoven University of Technology, 5600 MB Eindhoven, The Netherlands}
\affiliation{Max-Planck-Institut für Festkörperforschung, Heisenbergstraße 1, 70569 Stuttgart, Germany}
\author{Kathrin K{\"u}ster}
\author{Ulrich Starke}
\affiliation{Max-Planck-Institut für Festkörperforschung, Heisenbergstraße 1, 70569 Stuttgart, Germany}
\date{\today}

\begin{abstract}
Intercalation is an established technique for tailoring the electronic structure of epitaxial graphene. Moreover, it enables the synthesis of otherwise unstable two-dimensional (2D) layers of elements with unique physical properties compared to their bulk versions due to interfacial quantum confinement. In this work, we present uniformly Pb-intercalated quasi-freestanding monolayer graphene on SiC, which turns out to be essentially charge neutral with an unprecedented $p$-type carrier density of only $(5.5\pm2.5)\times10^9$ cm$^{-2}$. Probing the low-energy electronic structure throughout the entire first surface Brillouin zone by means of momentum microscopy, we clearly discern additional bands related to metallic 2D-Pb at the interface. Low-energy electron diffraction further reveals a $10\times10$ Moiré superperiodicity relative to graphene, counterparts of which cannot be directly identified in the available band structure data. Our experiments demonstrate 2D interlayer confinement and associated band structure formation of a heavy-element superconductor, paving the way towards strong spin-orbit coupling effects or even 2D superconductivity at the graphene/SiC interface.
\end{abstract}
\maketitle

\section{INTRODUCTION}
Graphene, the pioneer of the world of two-dimensional (2D) materials, has been a big subject undergoing intense investigations since its discovery almost 20 years ago \cite{novoselov2004,berger2004,novoselov2005,geim2007,novoselov2012}. During the last decade, the graphitization of SiC surfaces in argon atmosphere has matured into one of the major fabrication techniques for epitaxial graphene as it offers superb and uniform layer quality on a large scale and on a semiconducting substrate \cite{emtsev2009,kruskopf2016}. Furthermore, intercalation stands out as a versatile method to modify the properties of epitaxial graphene on SiC ('epigraphene') via the sandwiching of foreign atomic species at the interface. Many intercalants have been studied by now, each of them giving different doping and proximity effects in epigraphene \cite{riedl2009,mcchesney2010,walter2011,stoehr2016,link2019,rosenzweig2019,rosenzweig2020b,forti2020,rosenzweig2020,briggs2020}. At the same time, the technique offers a way to synthesize otherwise unstable 2D triangular lattices via their dimensional confinement at the graphene/SiC heterointerface. In turn, intriguing properties emerge such as superconductivity in intercalated 2D Ga \cite{briggs2020} or---in diametrical contrast to the respective bulk crystals---the opening of a semiconducting gap for the monolayer limit of Au \cite{forti2020} and Ag \cite{rosenzweig2020}.

Being a well-known superconductor even down to epitaxial ultrathin films \cite{zhang2010,brun2016} and also a heavy element with large spin-orbit coupling \cite{dil2008,yaji2010}, Pb has recently attracted a lot of attention as an intercalant for epigraphene, both experimentally \cite{yurtsever2016,chen2020,yang2021,hu2021,gruschwitz2021} and theoretically \cite{visikovskiy2018,wang2021,otheryang2021}. In fact, proximity spin-orbit coupling has been induced in Pb-intercalated graphene, yet only on metallic Ir and Pt substrates \cite{calleja2015,klimovskikh2017,otrokov2018}. On the other hand, proximity superconductivity has been reported in quasi-freestanding monolayer graphene on SiC decorated with Pb islands \cite{cherkez2018,paschke2020}. Regarding the very limited extent of this lateral proximity effect into graphene, the homogeneous intercalation of Pb underneath could be a pragmatic workaround as it might even give rise to vertical proximity coupling that spreads over the entire layer.

Previous experiments have been focussing on the induced doping level in epigraphene and the atomic structure upon Pb intercalation \cite{yurtsever2016,chen2020,yang2021,hu2021,gruschwitz2021}. Unfortunately, the interpretation was mostly complicated by the presence of multiple minority phases owing to partial, inhomogeneous intercalation in conjunction with a mixture of different graphene layer thicknesses on the pristine substrates. On the other hand, density functional theory suggests the stability of a triangular lattice monolayer of Pb in $1\times 1$ epitaxial relation to SiC \cite{visikovskiy2018,otheryang2021} but no experimental evidence could be provided so far. While the actual atomic and electronic structure of Pb confined between epigraphene and SiC remains an open question, it will be pivotal for the envisioned interlayer proximity couplings, thus reinforcing the need for a dedicated study. 

Here, we homogeneously intercalate Pb underneath the carbon buffer layer on SiC. Angle-resolved photoelectron spectroscopy reveals the characteristic $\pi$ bands of decoupled quasi-freestanding monolayer graphene, whose charge neutrality with a residual hole density on the order of $10^9$ cm$^{-2}$ holds great promise in terms of high carrier mobilities. Momentum microscopy readily provides access to the entire first surface Brillouin zone and uncovers additional electronic bands originating from intercalated Pb at the graphene/SiC heterointerface. We unambiguously demonstrate the metallic character of 2D confined Pb, rendering the system a hot candidate for superconductivity in the 2D metal itself or even in epigraphene via proximity coupling. While x-ray photoelectron spectroscopy and low-energy electron diffraction corroborate the complete intercalation scenario, the latter method reveals an additional $10\times10$ Moiré periodicity relative to graphene, which is yet difficult to reconcile with the available band structure data. Our results establish Pb-intercalated epigraphene as a promising platform to harvest intriguing quantum effects connected to the rich interlayer band structure of a 2D quantum-confined heavy metal.

\section{EXPERIMENT}\label{sec2}
\subsection{Sample preparation}
Nominally on-axis, single crystalline, $n$-doped 6H-SiC(0001) wafer pieces (SiCrystal GmbH) were used as substrates for graphene growth. In order to first remove residual polishing scratches and generate atomically flat terraces, the substrates were etched with molecular hydrogen at $1550$~$^{\circ}\mathrm{C}$ and near ambient pressure \cite{ramachandran1998,soubatch2005}. Heating to around 1465~$^{\circ}\mathrm{C}$ for $4$ min under $800$ mbar argon atmosphere (optimized parameters for the current setup) induces graphitization via Si sublimation \cite{emtsev2009}. In this way, several-$\mu$m-wide terraces develop, homogeneously covered with the $(6\sqrt{3}\times6\sqrt{3})\mathrm{R}30^\circ$ carbon buffer layer reconstruction. This so-called zerolayer graphene (ZLG) interacts covalently with the Si-terminated SiC substrate and does not yet share the electronic properties of a freestanding graphene monolayer \cite{emtsev2008,riedl2010}. Hydrogen etching and graphene growth were both performed \emph{ex situ} in an inductively heated reactor hosting a graphite susceptor.

The ZLG/SiC samples were subsequently transferred into ultrahigh vacuum (UHV) and degassed at $700$~$^{\circ}\mathrm{C}$ for $30$ min. Pb was evaporated onto ZLG at a rate of about 4~{\AA/min from a commercial Knudsen cell (OmniVac). During evaporation, the sample was kept at room temperature. In a first cycle, Pb was deposited for 10 min, followed by annealing at $200$~$^{\circ}\mathrm{C}$, $300$~$^{\circ}\mathrm{C}$, $400$~$^{\circ}\mathrm{C}$, $500$~$^{\circ}\mathrm{C}$, and $550$~$^{\circ}\mathrm{C}$ for $\approx 1$ h, respectively. As a result, Pb atoms migrate under the carbon buffer layer, saturating the Si dangling bonds and decoupling ZLG into quasi-freestanding monolayer graphene (QFMLG). However, only partial intercalation with large residual patches of ZLG was observed at this stage, while further annealing to $600$~$^{\circ}\mathrm{C}$ and $650$~$^{\circ}\mathrm{C}$ for $\approx 30$ min each already marked the onset of de-intercalation. A second cycle then included deposition and annealing steps identical to the first cycle, but only up to a temperature of $550$~$^{\circ}\mathrm{C}$. This rather long preparation protocol turned out to significantly enhance the homogeneity of Pb intercalation, also in comparison to previous efforts \cite{yurtsever2016,chen2020,yang2021,hu2021,gruschwitz2021}, ensuring at the same time the re-evaporation of excess Pb from the sample surface. All temperature measurements were performed with an infrared pyrometer (Impac IGA 140 series) assuming an emissivity of $0.63$.

\subsection{Characterization}
The quality of the intercalation scenario was monitored on the basis of low-energy electron diffraction (LEED), x-ray photoelectron spectroscopy (XPS) and angle-resolved photoelectron spectroscopy (ARPES). Photoelectron spectroscopy was carried out by means of an energy-filtered photoemission electron microscope (PEEM) operating at an extractor voltage of $12$ kV in both, real and $k$-space  (NanoESCA, Scienta Omicron GmbH). Single-shot-type ARPES constant energy cuts were acquired in the framework of momentum microscopy \cite{kroemker2008,tusche2015,tusche2019} from a $\approx 20\times 15$ $\mu\mathrm{m}^2$ spot on the sample surface as defined by an iris aperture. The spectrometer was operated at a nominal energy resolution of $0.1$ eV and a $k$-space field of view of 4.3~{\AA$^{-1}$} which covers the entire photoemission horizon up to the Fermi level for the employed photon energy of $21.22$ eV (non-monochromatized, unpolarized HeI $\alpha$ provided by a HIS 14 HD source, FOCUS GmbH). XPS spectra were acquired from a $\approx 60$ $\mu\mathrm{m}$ real-space field of view in energy-filtered PEEM mode at a spectrometer resolution of $0.4$ eV. Monochromatized Al K$\alpha$ radiation ($1486.6$ eV) was employed ($\mu$-FOCUS 350 monochromator, SPECS GmbH). Experiments were performed in UHV at room temperature and at a working pressure around $5\times10^{-10}$~mbar ($3\times10^{-9}$~mbar for XPS).

\section{RESULTS AND DISCUSSION}
\subsection{Low-energy electron diffraction}
Fig.~\ref{fig1}(a) presents the LEED pattern acquired after Pb intercalation for an incident beam energy of $70$ eV. A typical intercalation scenario is clearly evident from the intense first-order graphene diffraction (red) while the first order spots of ($1\times1$)-SiC as well as the $(6\sqrt{3}\times6\sqrt{3})\mathrm{R}30^\circ$ buffer layer reconstruction (green) are largely suppressed, cf.\ Refs.~\citenum{rosenzweig2019,link2019}. On the other hand, a Moiré pattern emerges around the first order spots of graphene as highlighted by the zoom-in on the 6-fold symmetrized LEED pattern in Fig.~\ref{fig1}(b). The corresponding spots do not coincide with the $6\sqrt{3}$ supercell grid of SiC and point towards a distinct ordered phase of intercalated Pb. We determine the Moiré periodicity as $10\times 10$ with respect to graphene via a 2D Lorentzian fit [Fig.~\ref{fig1}(c), corresponding line profile in Fig.~\ref{fig1}(d)]. This is consistent with earlier studies where a similar Moiré pattern was observed locally by scanning tunneling microscopy on partially Pb-intercalated epigraphene samples \cite{yurtsever2016,yang2021,hu2021,gruschwitz2021}.

\begin{figure}[t]
	\centering
    \includegraphics{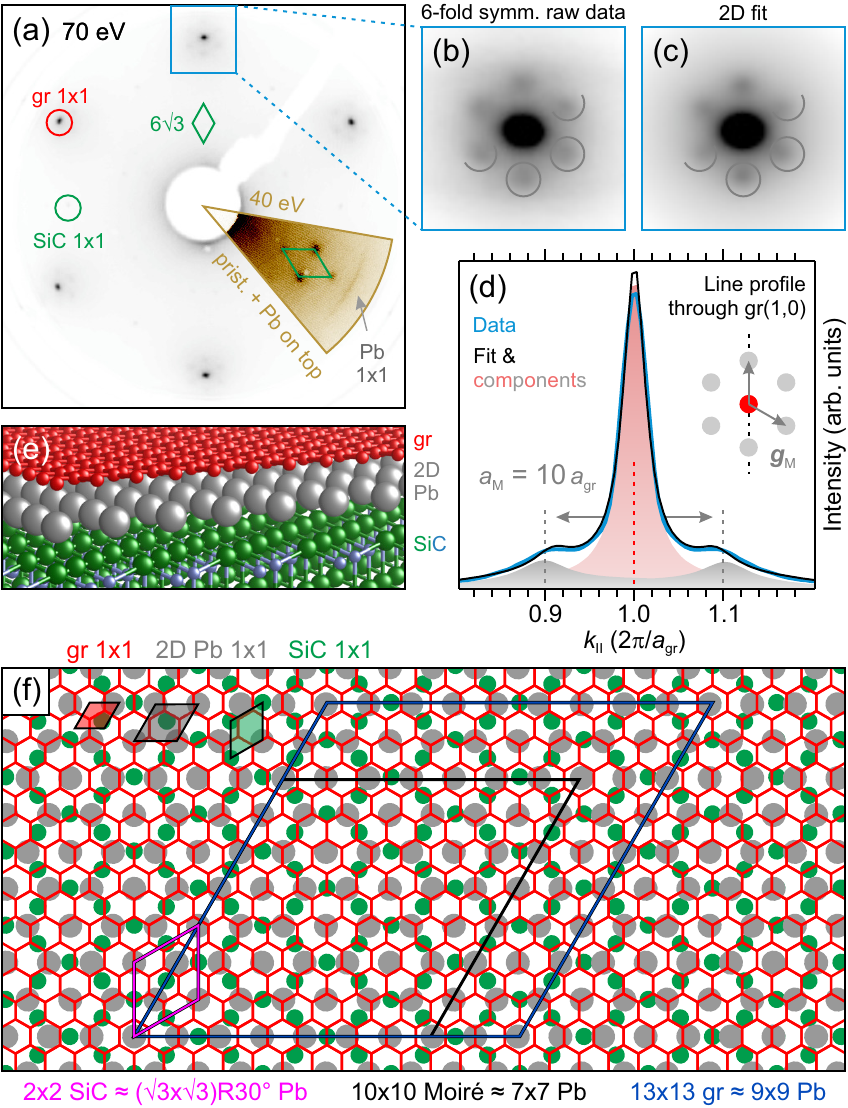}
	\caption{(a) LEED pattern of Pb-intercalated QFMLG at $70$ eV. Sector inset: $40$ eV LEED pattern of pristine ZLG with Pb islands on top. (b) Six-fold symmetrized raw data and (c) 2D lorentzian fit around the first order graphene spots, highlighting the induced Moiré periodicity (half of the corresponding spots encircled in grey). (d) Fitted vertical line profile through the symmetrized pattern of (b) with contributions from first-order graphene diffraction (red) and the $10\times10$ Moiré pattern (grey). (e) Side and (f) top views of the $13\times13$ graphene (red) on $\approx 9\times9$ Pb (grey) on $(6\sqrt{3}\times6\sqrt{3})\mathrm{R}30^\circ$ SiC (green) structure which approximates the $10\times10$ Moiré periodicity to better than $99$\%. Unit cells and superstructures are color-encoded in (f).}
	\label{fig1}
\end{figure}

Note that $10\times10$ graphene unit cells would perfectly coincide with a $7\times7$ supercell of intercalated Pb if the latter adopted the very same triangular lattice as the surface layer of Pb(111) islands deposited on top of epigraphene \cite{cherkez2018}.The sector inset of Fig.~\ref{fig1}(a) shows the LEED pattern of pristine ZLG with Pb on top acquired at $40$ eV. At this energy, the elongated spots originating in the direction of graphene due to first-order diffraction from rotationally disordered Pb(111) islands are clearly visible (grey arrow) \footnote{Note the different position of, e.g., the diamond-like spot arrangement of $6\sqrt{3}$-SiC [green lines in Fig.~\ref{fig1}(a)] on the LEED screen since the size of the Ewald sphere scales with electron energy.}. Additional LEED patterns are presented in the Supplemental Material, Fig.~S1 \cite{supplement}. It is however hard to imagine that 2D confined Pb between graphene and SiC [Fig.~\ref{fig1}(e)] retains the exact same lattice parameter of bulk-truncated Pb islands and behaves completely independent from the underlying SiC substrate. In fact, by imposing a tensile strain of only $\approx 1.2$\% to the isolated surface plane of a Pb(111) island, $(\sqrt{3}\times\sqrt{3})\mathrm{R}30^\circ$-Pb could be matched precisely to $2\times2$-SiC and an overall $13\times 13$ graphene on $9\times 9$ Pb on $(6\sqrt{3}\times 6\sqrt{3})\mathrm{R}30^\circ$ SiC supercell would result. This means that by adopting an intermediate strain level of less than $1$\%, 2D Pb could be made commensurate with $13\times 13$ graphene ($\approx 9\times 9$-Pb), $2\times 2$ SiC ($\approx (\sqrt{3}\times\sqrt{3})\mathrm{R}30^\circ$-Pb) and the $10\times10$ Moiré periodicity ($\approx 7\times 7$-Pb) to better than $99$\% on the scale of each single supercell. Fig.~\ref{fig1}(f) provides a schematic top view of the heterostacked lattices where the individual unit cells and superstructures are indicated. The composite system could then offer mechanisms to release the strain over a longer range (potentially involving internal reconstructions, buckling in 2D Pb) while ensuring commensurability of Pb with SiC and the Moiré superstructure on large-scale average.

However, the experimental LEED pattern lacks any spots that correspond directly to the modeled $\approx(\sqrt{3}\times\sqrt{3})\mathrm{R}30^\circ$-Pb on $2\times 2$-SiC superstructure. In particular, no first-order diffraction from the associated Pb lattice is observed throughout a $>200$ eV range of incident electron energy, in contrast to the case of epitaxial Pb islands on top of graphene [cf.\ inset of Fig.~\ref{fig1}(a)]. The model further opposes the prevalent tendency towards $1\times1$ epitaxial order of 2D heteroconfined metals relative to SiC \cite{forti2020,rosenzweig2020,briggs2020,yaji2019}. As will be discussed below, our ARPES data [Figs.\ \ref{fig2} and \ref{fig3}] rather support the latter scenario, which, on the other hand, cannot be readily reconciled with the $10\times 10$ Moiré pattern. It is therefore still unclear whether the observed Moiré phase is directly linked to the atomic arrangement of interfacial Pb and, if so, prevails throughout the entire surface or just relates to minority regions hosting a distinct atomic structure.

\subsection{Momentum microscopy of quasi-freestanding monolayer graphene}
Fig.~\ref{fig2}(a) shows a volumetric ARPES dataset of Pb-QFMLG acquired by means of momentum microscopy in steps of $50$ meV down to binding energies $E>5$ eV. The corresponding $\approx300$ $\mu\mathrm{m}^2$ real-space field of view is outlined in red in the underlying threshold-PEEM image. Adopting a characteristic crown shape, conically peaking at the six $\overline{\mathrm{K}}$ points of the hexagonal Brillouin zone (BZ) of graphene (dashed red overlay), the sharp ARPES intensity distribution indicates a completely decoupled, quasi-freestanding graphene monolayer very close to charge neutrality.

Extracted from the very same 3D dataset, the energy-momentum cut of Fig.~\ref{fig2}(b) displays the graphene $\pi$-band dispersion along the $\overline{\Gamma\mathrm{KM}\Gamma}$ wedge in the first BZ (cf.\ inset). Represented by the dashed red curves, a third nearest neighbor tight binding (3rd NN TB) model \cite{classen2020} has been fitted to the spectral maxima \footnote{Dirac-point energy $E_D=-9$ meV and Fermi velocity $v_F=6.85$ eV{\AA} have been fixed according to the detailed analysis of Fig.~\ref{fig2}(e) and (f).}. Besides matching the $\approx 2.85$ eV energy separation between the Dirac point $E_D$ at $\overline{\mathrm{K}}$ and the saddle-point Van Hove singularity $E_\mathrm{VHS}$ at $\overline{\mathrm{M}}$, the model describes the overall $\pi$-band course reasonably well throughout the entire probed energy-momentum range. This goes however at the expense of a physically intuitive size and ratio of the hopping parameters ($\gamma=3.58$, $-0.16$, and $0.18$ eV for first, second, and third NN hopping, respectively) which is similar to epigraphene at higher doping levels \cite{bostwick2007}.

\begin{figure*}[t]
	\centering
	\includegraphics{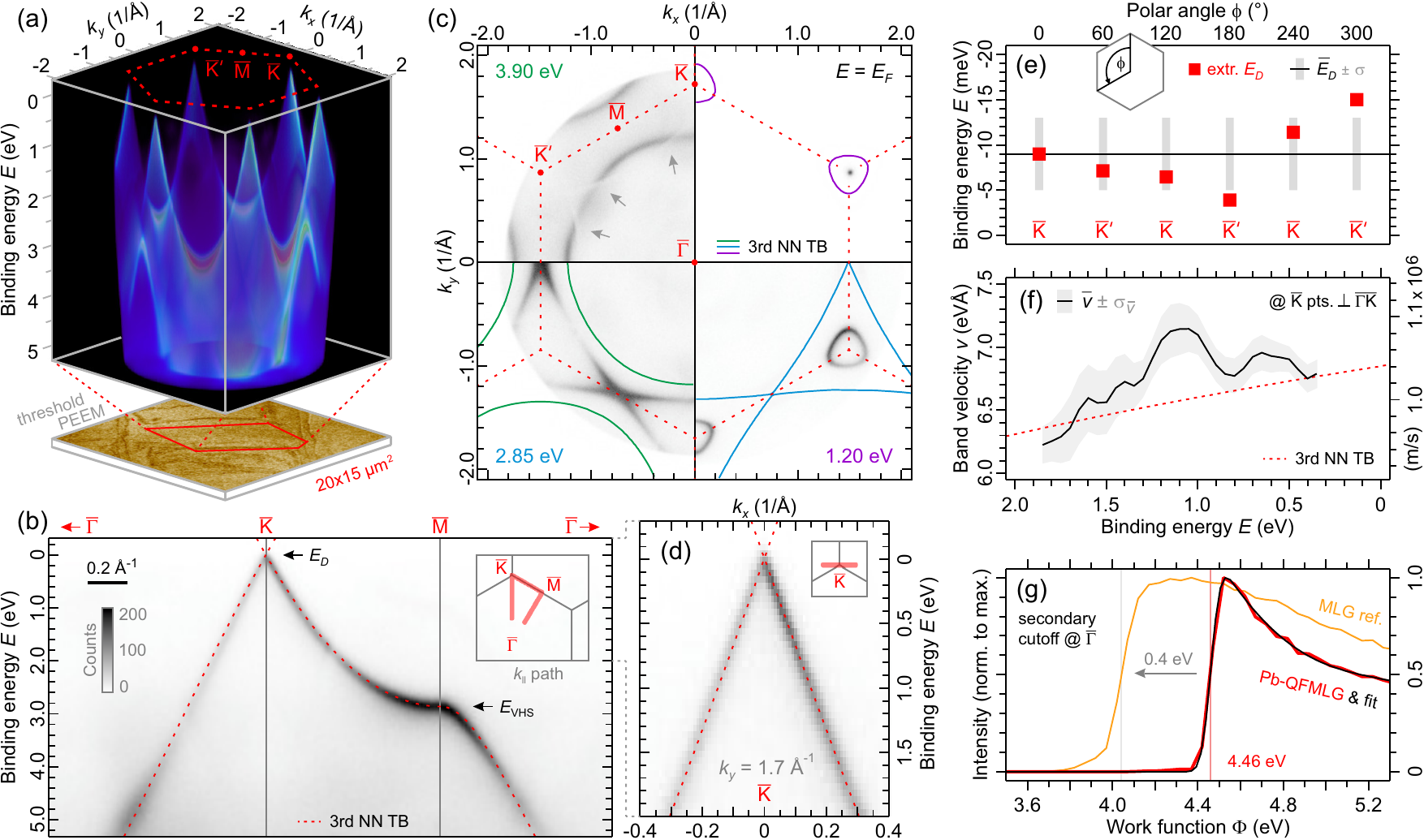}
	\caption{(a) Volume rendered 3D ARPES stack of Pb-QFMLG with the hexagonal BZ of graphene shown by dashed red lines. A region of about $20\times 15$ $\mu\mathrm{m}^2$ was probed in real space as indicated in the underlying threshold-PEEM image. (b) $\pi$ band dispersion extracted from (a) along the $\overline{\Gamma\mathrm{KM}\Gamma}$ wedge in the first BZ and fitted with a 3rd NN TB model \cite{classen2020} (dashed red curves). (c) Sequence of $k_x$-$k_y$ cuts at binding energies of $3.90$ eV, $2.85$ eV, $1.20$ eV, and at the Fermi level $E_F$. Each cut is clipped to a single $k$-space quadrant and the corresponding TB contour of QFMLG (solid green, blue, and purple curves, respectively) is overlayed to the counterclockwise adjacent quadrant for better visibility of the raw data. Spectral weight is partially suppressed due to a scattering-resonance effect (grey arrows) \cite{nazarov2013,krivenkov2017}. (d) Energy-momentum cut through the graphene Dirac cone at $\overline{\mathrm{K}}$, perpendicular to the $\overline{\Gamma\mathrm{K}}$ direction. Dashed red curves reproduce the TB fit according to panel (b). (e) A detailed determination of the Dirac-point binding energy $E_D$ for all six $\overline{\mathrm{K}}$ points (red squares) yields a mean value $\overline{E}_D=-9$ meV (black line) with standard deviation $\sigma=4$ meV (grey bars). (f) Averaged Dirac cone band velocity $\overline{v}$ perpendicular to $\overline{\Gamma\mathrm{K}}$ as a function of binding energy (black curve). The grey corridor gives the corresponding standard error of the mean $\sigma_{\overline{v}}$ and the dashed red curve represents the 3rd NN TB model. (g) The secondary spectral cutoff at $\overline\Gamma$ reveals a work function of $4.46$ eV for Pb-QFMLG (data: red, fit: black). A reference spectrum for MLG/SiC with a $\approx 0.4$ eV lower work function is also shown (yellow).}
	\label{fig2}
\end{figure*}

Fig.~\ref{fig2}(c) presents an overview of $k_x$-$k_y$ cuts at different binding energies relative to the Fermi level $E_F$ (as acquired directly by the NanoESCA momentum microscope). Each cut is clipped to a $k$-space quadrant and the respective 3rd NN TB contour as per the band structure fit of Fig.~\ref{fig2}(b) is overlayed to the counterclockwise adjacent quadrant (solid green, blue, and purple curves). The BZ of QFMLG is indicated by dashed red lines. An electron pocket around $\overline{\Gamma}$ (binding energy $E=3.90$ eV) transforms---via the $\overline{\mathrm{M}}$-point Van Hove singularity ($2.85$ eV)---into hole pockets around $\overline{\mathrm{K}}$ and $\overline{\mathrm{K}}$' ($1.20$ eV). The latter eventually converge into a singular, pointlike Fermi surface ($E=E_F$), reflecting the apparent charge-neutrality of Pb-QFMLG. For specific kinetic energies, excited photoelectrons can repeatedly scatter back and forth between graphene and its vacuum barrier in what is known as a scattering resonance \cite{nazarov2013,krivenkov2017}. Such a resonance entails a well-defined loss of spectral weight, dispersing steeply upwards from the vacuum level with near-unity effective mass and sharing the periodicity of the graphene host lattice. We thus attribute the sharp arcs of suppressed $\pi$-band intensity as highlighted by the grey arrows in the $k_x$-$k_y$ cut at $E=3.90$ eV [top left of Fig.~\ref{fig2}(c)] to repeated scattering resonances folding back into the first BZ \footnote{The scattering resonance centered around normal emission ($\overline{\Gamma}$) in the first BZ exceeds the available photoemission horizon}. Concomitantly, the sharp features confirm a clean surface of QFMLG down to the nanoscale without residual Pb clusters on top, whose presence would perturb the vacuum barrier and potentially destroy the scattering resonance condition completely via local rehybridization of graphene towards partial $sp^3$ character \cite{krivenkov2017}.

The energy-momentum cut of Fig.~\ref{fig2}(d) provides a near-$E_F$ closeup of the $\pi$-band dispersion through the $\overline{\mathrm{K}}$ point, perpendicular to the $\overline{\Gamma\mathrm{K}}$ direction ($k_x$ for fixed $k_y=1.7$ \AA$^{-1}$). A single, sharp Dirac cone converging in the absolute vicinity of $E_F$ demonstrates the uniform intercalation of Pb without any admixture of differential (minority) doping levels, thus representing a significant step beyond previous studies of Pb-intercalated epigraphene \cite{yurtsever2016,gruschwitz2021}. Dashed red curves overlayed to Fig.~\ref{fig2}(d) again trace the 3rd NN TB model as fitted to the wide-range $\overline{\Gamma\mathrm{KM}\Gamma}$ data of Fig.~\ref{fig2}(b). The model describes the experimental Dirac-cone dispersion remarkably well also in the direction perpendicular to $\overline{\Gamma\mathrm{K}}$, although the corresponding data have not been taken into account for the fit itself.

The apparent charge neutrality of Pb-QFMLG can still be quantified more precisely. Fig.~\ref{fig2}(e) shows the Dirac-point energy $E_D$, determined individually for all six $\overline{\mathrm{K}}$\textsuperscript{(}'\textsuperscript{)} points via the crossing of linear fits within $0.2$--$1.0$ eV down from $E_F$ to the extracted dispersion in the direction perpendicular to $\overline{\Gamma\mathrm{K}}$ \footnote{Values of $E_D$ have also been corrected for meV-sized variations in Fermi-edge position with momentum, resulting from the non-isochromaticity \cite{tusche2019} and putative inhomogeneities in the electronic lens system of the spectrometer}. All $\overline{\mathrm{K}}$\textsuperscript{(}'\textsuperscript{)} points reveal very small $p$-type doping with a Dirac crossing slightly above the Fermi level ($E_D<0$). The mean value turns out as $\overline{E}_D=-9$ meV with standard deviation $\sigma=4$ meV. Using the well-known relation $n=E_D^2/(\pi v_F^2)$ where the Fermi velocity $v_F$ is given in units of eV{\AA} [cf.\ Fig.~\ref{fig2}(f)], we determine a residual hole density for $p$-type doped Pb-QFMLG of only $n=(5.5\pm2.5)\times10^9$ cm$^{-2}$. Corresponding only to a single charge carrier per $\approx700,000$ C atoms, this vividly highlights the essential charge neutrality of Pb-QFMLG. The latter is interesting in terms of the overall limited carrier mobilities of epigraphene, which have often been ascribed to interactions with its substrate as evidenced by, e.g., finite doping \cite{emtsev2009,sinterhauf2020,aprojanz2020}. Note that epigraphene is generally doped via (i) the spontaneous polarization for hexagonal SiC \cite{ristein2012}, (ii) excess-charge transfer for $n$-type SiC \cite{mammadov2014}, and (iii) element-specific charge transfer from intercalated layers \cite{riedl2009,mcchesney2010,walter2011,stoehr2016,link2019,rosenzweig2019,rosenzweig2020b,forti2020,rosenzweig2020,briggs2020}.
Future studies will thus have to clarify whether the charge neutrality of Pb-QFMLG results from perfect screening of all substrate contributions while simply no charge transfer takes place from interfacial Pb onto graphene or is merely due to coincidential compensation of all doping mechanisms (i)--(iii). In the latter case, the depletion of $n$-type SiC alone can be expected to contribute carrier densities on the order of $10^{12}$ cm$^{-2}$ \cite{mammadov2014}. A suppression of doping mechanism (ii), as anticipated well below the bulk dopants' freeze-out temperature or for semi-insulating SiC, should therefore entail a substantial upshift of $E_D$ and thus help to elucidate the origin of charge neutrality in Pb-QFMLG.

Fig.~\ref{fig2}(f) shows the mean absolute band velocity $\overline{v}$ of the Dirac cone perpendicular to $\overline{\Gamma\mathrm{K}}$ as a function of binding energy, given by the smoothed derivative $\vert dE/dk\vert$ of the averaged experimental dispersion over all six BZ corners, i.e.\ 12 $\pi$-band branches (solid black curve). The grey corridor indicates the standard error of the mean $\sigma_{\overline{v}}$ as per the individual datasets and the dashed red curve represents once again the 3rd NN TB model [cf.\ Figs.~\ref{fig2}(b)--(d)]. Towards $E_F$ the inferred value of $\overline{v}= 6.9\pm0.1$ eV{\AA} turns out consistent with the well-established Fermi velocity of charge-neutral graphene \cite{castroneto2009}. However, $\overline{v}$ is found to peak out from the strictly monotonic TB curve at binding energies of $\approx 0.6$ eV and, even more prominent, $\approx 1.1$ eV. Such renormalizations differ from the gradual changes in band velocity previously observed for graphene in response to its dielectric environment \cite{siegel2011,hwang2012}. They could instead point towards hybridization with potential interlayer electronic states of Pb, although the latter cannot be readily discerned from the dominant $\pi$ bands of QFMLG (at least for the available photon energy). Note that such hybridization effects have previously been linked to enhanced spin splitting in the Dirac cone of other intercalated graphene systems \cite{marchenko2012,marchenko2016}. Similar effects might hence occur also in Pb-QFMLG, suggesting future studies by means of spin-resolved ARPES.

In order to compress the entire photoelectron hemisphere into the narrow angular acceptance cone of the NanoESCA momentum microscope, the sample is by default subject to an extractor voltage. This provides immediate access to the secondary spectral cutoff where the slowest photoelectrons pile up, barely overcoming the sample work function at vanishing parallel momentum. Fig.~\ref{fig2}(g) shows the corresponding normal emission spectrum for Pb-QFMLG (red curve) fitted with the product of an error function and an exponential decay (black curve). The horizontal axis is given in kinetic energy relative to $E_F$ so that the work function $\Phi=4.46\pm0.02$ eV of Pb-QFMLG can be directly read off from the secondary cutoff. The latter is just broadened to the $0.1$ eV resolution limit of the spectrometer, suggesting the absence of domains with differing work functions and reinforcing the homogeneity of the intercalated sample. At the same time, the work function of Pb-QFMLG turns out $\approx 0.4$ eV higher as compared to epitaxial monolayer graphene (MLG) on SiC [cf.\ yellow reference spectrum in Fig.~\ref{fig2}(g)]. This work function difference is well consistent with the Dirac-point energy of Pb-QFMLG (charge neutral, $E_D\approx E_F$) relative to MLG (moderately $n$ doped, $E_D\approx 0.4$ eV) \cite{riedl2010}. The work function of Pb-QFMLG thus appears to be governed by a mere Fermi-level shift while the vacuum level itself remains unperturbed. As already indicated by the presence of sharp scattering resonances [Fig.~\ref{fig2}(c)], the rigid behavior of $\Phi$ with $E_D$ corroborates a proper carbon-honeycomb surface termination without residual Pb left behind on top.

\subsection{Interlayer electronic structure of Pb}
Based on the LEED Moiré pattern which hints towards an ordered phase of intercalated Pb (cf.\ Fig.~\ref{fig1}) and the experience from other intercalants \cite{forti2020,rosenzweig2020,briggs2020}, interlayer band structure formation in 2D confined Pb is well conceivable and warrants specific focus when studying the electronic structure of Pb-QFMLG. However, previous ARPES measurements of (partially) Pb-intercalated epigraphene are entirely limited to single energy-momentum cuts through the $\overline{\mathrm{K}}$ point of graphene \cite{yurtsever2016,gruschwitz2021} where the dominant $\pi$ bands easily overshine any potential interlayer electronic states associated with intercalated Pb. In this regard, momentum microscopy readily allows for a more complete band structure survey of the system as it covers the entire BZ in a single-shot-type measurement.

\begin{figure*}[t]
	\centering
    \includegraphics{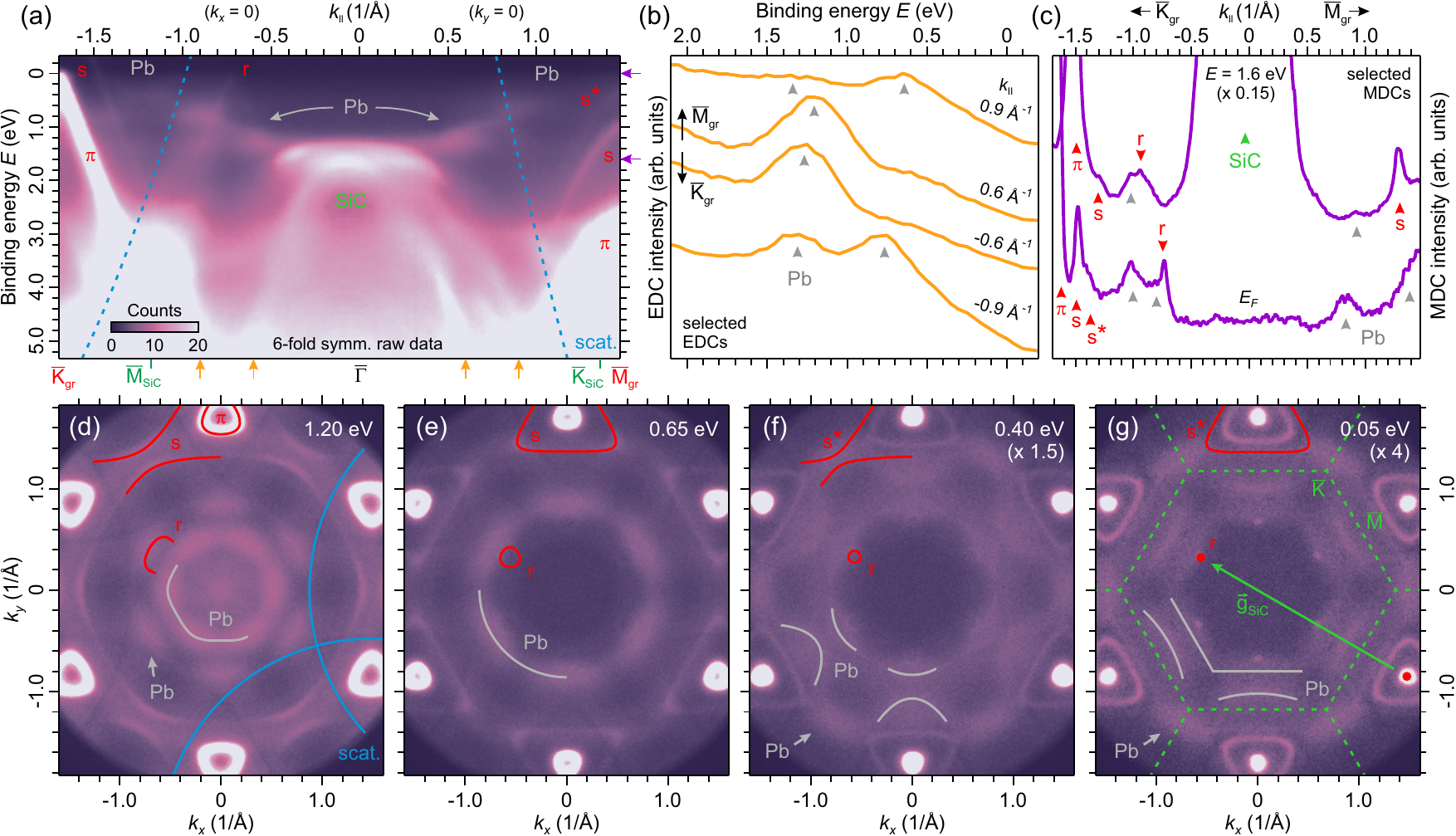}
	\caption{(a) ARPES cut along the $\overline{\mathrm{K}\Gamma\mathrm{M}}$ direction of graphene (equivalent to $\overline{\mathrm{M}\Gamma\mathrm{K}}$ of SiC) extracted from the dataset of Fig.~\ref{fig2} after 6-fold symmetrization to improve the signal-to-noise ratio of weaker spectral features. (b) Corresponding EDCs and (c) MDCs taken at selected momenta and binding energies [cf.\ colored arrows next to panel (a)]. (d)--(g) Series of $k_x$-$k_y$ cuts at $1.20$, $0.65$, $0.40$, and $0.05$ eV below the Fermi level $E_F$ with overlayed guides to the eye. The surface BZ of SiC is indicated in (g) by dashed green lines. $\pi$-band satellites (s and s$^*$, red) due to HeI $\beta$, $\gamma$ as well as Dirac cone replicas (r, red) backfolded via the surface reciprocal lattice vectors $\vec{g}$ of SiC are marked. Additional bands can be assigned to bulk SiC [(a) and (c), green] and intercalated Pb (grey). The free-electron dispersion of the scattering resonances is also indicated [(a) and (d), blue]. Counts in panels (f) and (g) have been multiplied by factors of $1.5$ and $4$, respectively.}
	\label{fig3}
\end{figure*}

Fig.~\ref{fig3}(a) shows an energy-momentum cut along the $\overline{\mathrm{K}\Gamma\mathrm{M}}$ path of graphene (corresponding to the $\overline{\mathrm{M}\Gamma\mathrm{K}}$ direction of SiC), whose color scale has been optimized so as to improve the visibility of any weaker spectral features. In order to also enhance their signal-to-noise ratio, the ARPES cut has only been extracted after 6-fold rotational symmetrization of the parent dataset of Fig.~\ref{fig2} \footnote{This procedure should be consistent with any 2D hexagonal arrangement of intercalated Pb, i.e., the LEED-based model of Fig.~\ref{fig1} or $1\times1$ epitaxial order as previously reported for other intercalants. At the same time, the threefold symmetry of bulk SiC is sacrificed.}. While the dominant $\pi$ bands of QFMLG oversaturate the applied color scale by up to a factor of ten, their weak HeI $\beta$ and $\gamma$ satellites can now be discerned, displaced by the difference in photon energy of $1.87$ eV (s) and $2.52$ eV (s$^*$), respectively. In addition, a replicated Dirac cone is made visible along the $\overline{\Gamma\mathrm{K}}$ direction, centered at $k_\parallel\approx -0.7$ \AA$^{-1}$ (labelled r). It can be attributed to backfolding of the opposite $\overline{\mathrm{K}}$-point region ($k_\parallel>0)$ via a surface reciprocal lattice vector of SiC as discussed below. The hole-like bands centered around normal emission are bulk states of SiC. Their maximum at $E\approx 1.3$ eV turns out consistent with the expected binding energy of the global valence band maximum of SiC (located at $\Gamma$ in the bulk BZ) when taking into account the upward band bending as measured by XPS (cf.\ Fig.~\ref{fig4}).

Beyond these distinct features, the ARPES cut of Fig.~\ref{fig3}(a) uncovers additional dispersive states which can neither be assigned to graphene nor SiC and must therefore be related to intercalated Pb. A corresponding band starts out electron-like from $E\approx 1.3$ eV at $\overline{\Gamma}$ where it is still eclipsed by the SiC valence band maximum. It initially demonstrates a relatively flat upward dispersion in both orthogonal momentum directions until splitting up into an upper and lower branch around $\vert k_\parallel\vert \approx 0.7$ \AA$^{-1}$. The evolution of a single Pb-related spectral peak ($k_\parallel=\pm 0.6$ \AA$^{-1}$) into two distinct peaks ($k_\parallel=\pm 0.9$ \AA$^{-1}$) is specifically highlighted by the corresponding energy distribution curves (EDCs) in Fig.~\ref{fig3}(b). While the upper branch retains an electron-like dispersion also beyond the splitting region, the lower branch rather adopts a negative curvature and appears to turn into a steep downward dispersion. However, tracing the Pb-related bands becomes increasingly difficult for higher values of parallel momentum ($k_\parallel\gtrsim 1$ \AA$^{-1}$) as their spectral weight largely diminishes and in particular the lower branch can no longer be discerned. Just like the spectral weight loss of the upper branch upon approaching the Fermi level, this could be linked to a change in orbital character as Pb offers both, $6s$ and $p$ valence electrons. Future studies with varying photon energy and polarization are therefore highly desirable in order to selectively enhance these weak Pb-related signatures close to $E_F$ and further clarify their precise dispersion.

Fig.~\ref{fig3}(c) shows two momentum distribution curves (MDCs) extracted from the ARPES cut in Fig.~\ref{fig3}(a) at a binding energy of $1.6$ eV (scaled to $15$\% intensity) and right at $E_F$. Colored markers indicate contributions of graphene's $\pi$ bands including their satellites (s, s$^*$) and replicas (r) as well as SiC bulk states. Additional MDC peaks (grey markers) are due to intercalated Pb and their presence at $E_F$ (bottom MDC) clearly confirms the metallic nature retained by the intercalant. This behavior is in complete contrast to the group $11$ noble metals which develop a global semiconducting gap when intercalated as monolayers beneath epigraphene \cite{forti2020,rosenzweig2020}. The metallic character of intercalated Pb at the graphene/SiC heterointerface could for instance be related to a different atomic arrangement (cf.\ Fig.~\ref{fig1}) or the distinct valence electron configuration of Pb comprising also $p$ electrons. In any case, as 2D confined Pb on SiC remains metallic, it is well conceivable that also the superconducting properties are inherited from the bulk crystal---just like for epitaxial ultrathin films of Pb on Si(111) \cite{zhang2010,brun2016}. Our experimental findings therefore hold great promise and motivate future experiments in view of a potential superconducting transition in Pb-QFMLG.

A series of 6-fold symmetrized $k_x$-$k_y$ cuts at different binding energies is shown in Fig.~\ref{fig3}(d)--(g). At $1.20$ eV below $E_F$ [Fig.~\ref{fig3}(d)], the electron pocket centered at $\overline{\Gamma}$ and demonstrating slight hexagonal warping corresponds to the flat bottom of the Pb-related band. Likewise, the adjacent spectral weight around $k_\parallel= 0.8$ \AA$^{-1}$ towards $\overline{\mathrm{K}}_\mathrm{gr}$ ($\overline{\mathrm{M}}_\mathrm{SiC}$) can be assigned to the lower, split-off Pb branch which acquires a hole-like dispersion. In addition, replicated Dirac cones (r) also contribute to the concerned regions in $k$-space as discussed below. The Pb-related features appear to be largely suppressed outside the curved hexagonal corridor enclosed by the giant scattering resonance arcs (blue curves).

For a binding energy of $0.65$ eV [Fig.~\ref{fig3}(e)], a single electron pocket is found to encircle $\overline{\Gamma}$ at a radius of $\approx 0.9$ \AA$^{-1}$ and with enhanced intensity in the direction of $\overline{\mathrm{K}}_\mathrm{gr}$. It corresponds to the upper branch of the Pb-related band in the energy-momentum cut of Fig.~\ref{fig3}(a) (for a corresponding zoom-in with enhanced contrast see Supplemental Material, Fig.~S2 \cite{supplement}). Additional features or repeated contours due to intercalated Pb are however not yet readily discernible at higher values of $k_\parallel$ in Fig.~\ref{fig3}(d) and (e). This is further complicated by the dominant spectral weight of graphene's $\pi$ bands and also their HeI $\beta$ satellites (red curves) towards the photoemission horizon.

At $0.4$ eV below $E_F$ [Fig.~\ref{fig3}(f)] the $k_x$-$k_y$ cut turns out further enriched. Additional concave segments now appear towards $\overline{\mathrm{K}}_\mathrm{gr}$, almost touching the inner contour at about $1$ \AA$^{-1}$. Along the $\overline{\mathrm{KMK}}$' line of graphene, the ARPES intensity appears to be also enhanced (centered around $\overline{\mathrm{M}}_\mathrm{gr}$). This has to be attributed largely to the intercalant as the minimal satellite-photon flux of HeI $\gamma$ alone will not be sufficient to probe the saddle-point-region of QFMLG (s$^*$, red curves) at an adequate intensity.

In direct vicinity of the Fermi level, the intercalant-related features appear somewhat sharper [Fig.~\ref{fig3}(g)]. This is however accompanied by an overall intensity loss and the counts in this panel have been quadrupled for enhanced visualization. An inner hexagonal contour is now found surrounded by additional segments at $k_\parallel\approx 1$ \AA$^{-1}$, running almost parallel and close to the border of the first SiC surface BZ (the latter indicated by the dashed green overlay). These distinct contours correspond to the two Fermi level crossings marked along $\overline{\Gamma\mathrm{K}}_\mathrm{gr}$ in the bottom MDC of Fig.~\ref{fig3}(c). They are also highlighted in the energy-momentum cut of Fig.~S2 in the Supplemental Material \cite{supplement}. Note that the slightly curved segments close to the SiC surface BZ border seem to develop out of the broader, concave contours visible in Fig.~\ref{fig3}(f) and actually move towards $\overline{\Gamma}$ when tracing their dispersion coming from higher binding energies. While the energy-momentum cut of Fig.~\ref{fig3}(a) \cite{supplement} can be interpreted accordingly in the vicinity of $\overline{\mathrm{M}}_\mathrm{SiC}$, repeated counterparts of the Pb band that originates in the first BZ cannot be readily identified at higher values of $k_\parallel$ as the dominant spectral weight of the graphene $\pi$ bands and even their satellites sets in. Likewise, this holds for the $k_x$-$k_y$ cut of Fig.~\ref{fig3}(g) where potential counterparts of the Pb-related arcs ($k_\parallel\approx 1$ \AA$^{-1}$) are easily overshadowed by satellite $\pi$-pockets of QFMLG (s$^*$, red curve). Therefore, the actual BZ of the intercalant is hitherto difficult to determine based solely on the accessible dispersion of extrinsic, Pb-related bands. Note that the enhanced spectral weight around $\overline{\mathrm{M}}_\mathrm{gr}$ which is attributed to Pb persists in Fig.~\ref{fig3}(g) while the HeI $\gamma$ satellite pockets of graphene (s$^*$, red curve) converge towards $\overline{\mathrm{K}}_\mathrm{gr}$.

As discussed above, the actual BZ geometry of intercalated Pb is difficult to deduce from the available data of Fig.~\ref{fig3} due to the lack (weak intensity and unclear dispersion, respectively) of periodically repeated band features and their corresponding constant energy contours. However, weak Dirac cone replicas (r) appear at $k_\parallel\approx 0.7$ \AA$^{-1}$ along $\overline{\Gamma\mathrm{K}}_\mathrm{gr}$ as highlighted by red curves in Fig.~\ref{fig3}(d)--(f). These replicas are generated by backfolding of the opposite BZ corners of graphene via the surface reciprocal lattice vectors $\vec{g}$ of SiC as indicated in Fig.~\ref{fig3}(g). As QFMLG is electronically decoupled from the SiC substrate via intercalation, its photoelectrons should predominantly feel the periodicity of the underlying intercalant in terms of final state photoelectron diffraction. Hence, the QFMLG replicas generated by $\vec{g}_\mathrm{SiC}$ suggest a $1\times 1$ epitaxial arrangement of interfacially confined Pb relative to SiC, that is, identical surface BZs as previously observed for other intercalants \cite{forti2020,rosenzweig2020,briggs2020}. In contrast, our ARPES data do not reflect the $(\sqrt{3}\times\sqrt{3})\mathrm{R}30^\circ$-Pb on $2\times 2$-SiC superstructure \footnote{A corresponding first BZ of Pb would be oriented along graphene and have its $\overline{\mathrm{K}}$ points coinciding with $\overline{\mathrm{M}}_\mathrm{SiC}$} as deduced from the $10\times 10$ Moiré superperiodicity in LEED (cf.\ Fig.~\ref{fig1}). The same holds for the Moiré periodicity itself, counterparts of which can not be identified in the momentum microscopy data of Fig.~\ref{fig3}. We note, however, that the kinetic energy of the photoexcited electrons in the present ARPES measurement is substantially lower as compared to the $70$ eV LEED pattern of Fig.~\ref{fig1}. Moiré replicas might hence reemerge when probing the electronic structure at higher photon energies. To this end, synchrotron-based ARPES at tunable photon energy would be desirable, not least in view of optimizing the cross section for the intriguing interlayer bands of Pb.

\subsection{X-ray photoelectron spectroscopy}
\begin{figure*}[t]
	\centering
    \includegraphics{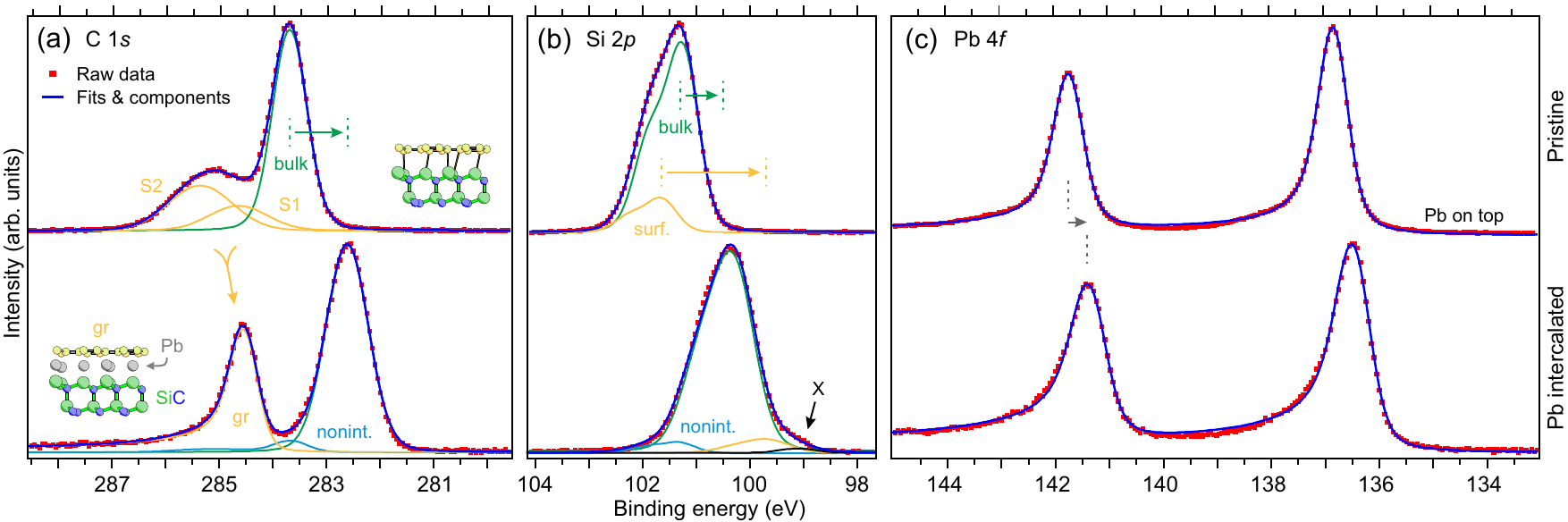}
	\caption{Fitted XPS spectra of (a) C $1s$, (b) Si $2p$, and (c) Pb $4f$ for pristine ZLG (top) and Pb-intercalated QFMLG (bottom). The top spectrum in (c) corresponds to nominally $\approx 8$ nm of Pb deposited onto pristine ZLG. Insets in (a) represent schematic ball-and-stick side views of the system and individual fit components are explained in the text.}
	\label{fig4}
\end{figure*}
We finally discuss the evolution of selected XPS core-level spectra, acquired from a $3600$ $\mu\mathrm{m}^2$ area before and after the intercalation of Pb \footnote{Spectra were fitted using the free software AAnalyzer\textregistered, \url{https://rdataa.com/download}. Unless explicitly mentioned otherwise, Voigt peaks with a fixed Lorentzian width of $0.1$ eV were employed.}. Fig.~\ref{fig4}(a) displays the well-established C $1s$ spectrum of pristine ZLG (top) with an intense bulk component (green curve) at a binding energy of $283.6$ eV and two surface components S1, S2 (yellow curves) at $284.6$ eV and $285.3$ eV, respectively. Recall that S1 represents the carbon atoms of ZLG covalently bonded to the SiC substrate while S2 accounts for those not interacting with the substrate (cf.\ ball-and-stick inset) \cite{riedl2010}. As about every third carbon atom of ZLG binds to the substrate \cite{riedl2010} the area ratio S2:S1 has been fixed at 2:1 for the fit. After the intercalation of Pb, it is now the intercalant atoms that saturate the Si dangling bonds of the substrate and the covalent interaction with the graphene layer is lifted. Consequently, S1 and S2 are replaced by a single QFMLG peak at $284.5$ eV (bottom spectrum) that has been fitted with an asymmetric Doniach-Sunjic line shape. The corresponding asymmetry parameter $\alpha$ is expected to scale with the carrier density of graphene \cite{sernelius2015,link2019,rosenzweig2019}. However, despite the charge neutrality of Pb-QFMLG, we find $\alpha=0.18$ which already exceeds the values reported for other epigraphene systems at substantially higher doping levels \cite{riedl2009,riedl2010}. Largely impeded by the pointlike Fermi surface of Pb-QFMLG itself (cf.\ Fig.~\ref{fig2}), the asymmetry could in turn be explained by the coupling to elementary excitations around the Fermi level in the metallic Pb layer (cf.\ Fig.~\ref{fig3}). Pb intercalation further shifts the bulk C $1s$ component of SiC by about $1$ eV to lower binding energies. Such behavior is characteristic for many different intercalants and can be attributed to the associated change in surface band bending at the SiC/graphene interface \cite{riedl2009,rosenzweig2019,rosenzweig2020}. Note that we have introduced the spectrum of pristine ZLG as an additional fit component (light blue curve), confirming a ratio $<5$\% of residual non-intercalated regions on the sample surface. This highlights the enhanced completeness of Pb intercalation, yielding a more distinct spectral shape as compared to earlier XPS studies \cite{yurtsever2016,yang2021,gruschwitz2021}.

The Si $2p$ spectrum for pristine ZLG/SiC is shown in the top part of Fig.~\ref{fig4}(b). It consists of a bulk doublet (green curve) at $101.3$ eV and a surface doublet (yellow curve) shifted by $0.35$ eV to higher binding energies due to covalent bonding of the topmost Si atoms to the carbon buffer layer. For both doublets, spin-orbit splitting and branching ratio have been fixed at $0.62$ eV and $0.5$, respectively. Upon Pb intercalation (bottom spectrum), the Si $2p$ bulk component undergoes the same $1$ eV shift to lower binding energies as observed for bulk C $1s$, driven by an upward band bending at the substrate/intercalant interface. The surface Si component concurrently shifts by almost $2$ eV in the same direction, ending up on the low binding energy side of its bulk counterpart. This highlights the modified interfacial bonding characteristics in the intercalated system now that the surface Si dangling bonds are no longer saturated by the carbon buffer layer but by Pb. Apart from the small contribution of non-intercalated residuals (light blue curve) which slightly broadens the spectrum towards higher binding energies, note the clear shoulder emerging on the low binding energy side as indicated by the black arrow. The latter can only be properly captured by introducing a dedicated fit component (X, black curve) at a binding energy of $99.1$ eV. While it could be due to substrate defects induced during the lengthy intercalation procedure (see Sec.~\ref{sec2}), no corresponding counterpart can readily be identified in the C $1s$ spectrum of Fig.~\ref{fig4}(a). The feature therefore requires further clarification beyond the scope of the present work.

A Pb $4f$ reference spectrum is displayed in the top part of Fig.~\ref{fig4}(c), acquired for nominally $\approx 8$ nm of Pb deposited onto pristine ZLG. The metallic character becomes directly evident from the asymmetric line shape, fitted by a single Doniach-Sunjic doublet. The binding energy of $136.9$ eV for the $4f_{7/2}$ level and the spin-orbit splitting of $4.86$ eV turn out fully consistent with the literature values for bulk Pb \cite{xpshandbook}. After intercalation (bottom), the Pb $4f$ doublet retains its asymmetric shape and thereby confirms the metallic nature of 2D confined Pb at the epigraphene/SiC interface as unambiguously demonstrated by our ARPES results (cf.\ Fig.~\ref{fig3}). On the other hand, the doublet shifts by $0.4$ eV to lower binding energies as compared to the bulk reference spectrum. This reflects the different chemical environment of intercalated Pb and its interaction (bonding) with the Si-terminated substrate, in line with the displaced Si $2p$ surface doublet [Fig.~\ref{fig4}(b)]. Distinct components of intercalated Pb cannot be resolved at this point and we therefore conclude that all intercalated atoms seem to interact with the substrate in a comparable way, thereby rendering a multilayer configuration of interfacial Pb unlikely. Likewise, the absence of an additional Pb $4f$ doublet at higher binding energy (cf.\ bulk reference spectrum) confirms that there is no residual Pb left on top of QFMLG, fully consistent with the ARPES results in Fig.~\ref{fig2}.

\section{CONCLUSION AND OUTLOOK}
In summary, we report the homogeneous intercalation of Pb underneath a single layer of epitaxial graphene on SiC. Momentum microscopy across the entire surface Brillouin zone is used to probe the sharp $\pi$ bands of the resulting quasi-freestanding graphene monolayer which hosts a negligible hole density of only $5.5\times10^9$ cm$^{-2}$. Additional dispersive bands due to 2D quantum-confined Pb at the graphene/SiC heterointerface can also be identified and the metallic character retained by Pb upon intercalation is clearly demonstrated. Our band structure data hint towards $1\times1$ epitaxial order of intercalated Pb relative to SiC. In addition, a $10\times 10$ Moiré superperiodicity with respect to graphene emerges in low-energy electron diffraction, whose counterparts could not yet be identified in the electronic structure.

On the one hand, our findings hold great promise in view of enhanced carrier mobilities in the epigraphene layer which is rendered charge neutral on the large scale via the homogeneous intercalation of Pb. On the other hand, the demonstrated large-area synthesis of quantum-confined Pb with a metallic band structure opens a route towards 2D superconductivity and strong spin-orbit coupling at the modified epigraphene/SiC interface. These anticipated properties might not even be limited to the intercalant itself but could eventually propagate onto overhead graphene through vertical proximity coupling---boosted by the enhanced homogeneity of the intercalation scenario reported herein. Our study identifies Pb-intercalated epigraphene as a viable candidate to harvest intriguing quantum properties of a 2D-confined heavy-element superconductor. Providing the desired sample homogeneity and a proof of emerging interlayer electronic states, our results form a solid basis for, e.g., low-temperature transport measurements as well as detailed band structure studies by means of synchrotron-based spin- and angle-resolved photoelectron spectroscopy at maximal resolution.

\begin{acknowledgments}
The authors would like to thank Hrag Karakachian for most valuable discussions. Funding by the Deutsche Forschungsgemeinschaft (DFG) through Sta315/9-1 and FOR 5242 is gratefully acknowledged. O.B.\ was supported by the Erasmus$+$ Exchange Programme of the European Union.
\end{acknowledgments}

\newpage
\onecolumngrid
\begin{center}
\textbf{\large\emph{Supplemental Material for}}\\[0.1cm]
\textbf{\large Momentum microscopy of Pb-intercalated graphene on SiC: charge neutrality and electronic structure of interfacial Pb}
  \end{center}
  \setcounter{figure}{0}
  \renewcommand{\thefigure}{S\arabic{figure}}
  \vspace{\fill}
  \begin{figure*}[h]
	\centering
	\includegraphics[scale=0.95]{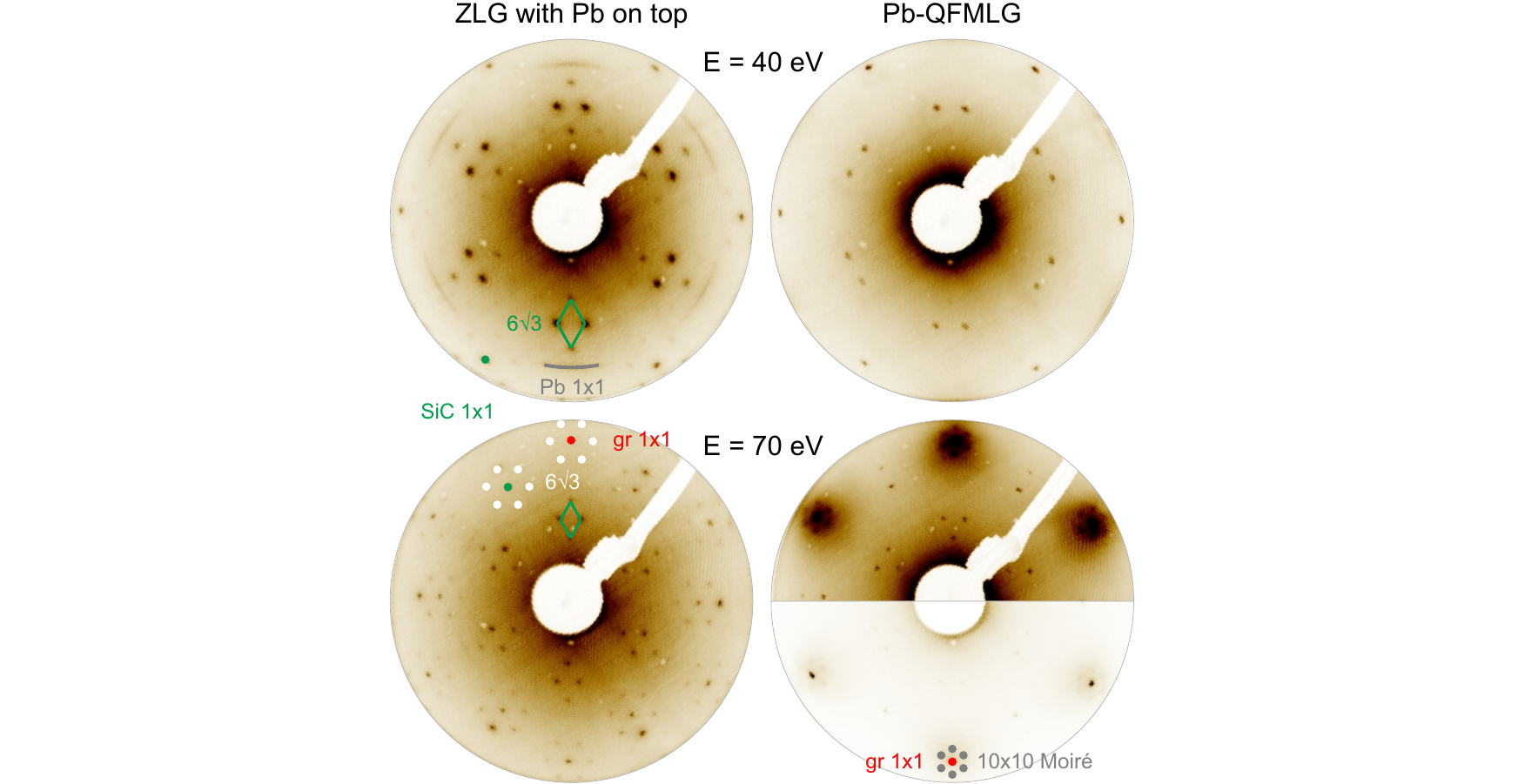}
	\caption{LEED patterns for pristine ZLG with Pb islands deposited on top (left) and for QFMLG obtained via Pb intercalation (right) at $40$ eV (top) and $70$ eV (bottom). Diffraction spots are indicated by colored markers. The same color scale has been applied to all patterns except for the bottom half of the $70$ eV pattern of Pb-intercalated QFMLG where the upper threshold has been increased by a factor of $10$ (corresponding to the strongly increased intensity of the first order graphene spots with respect to those of SiC or the $6\sqrt{3}$ supercell due to the decoupling of graphene from the substrate).}
	\label{figS1}
\end{figure*}
\vspace{\fill}
\begin{figure*}[h]
	\centering
	\includegraphics{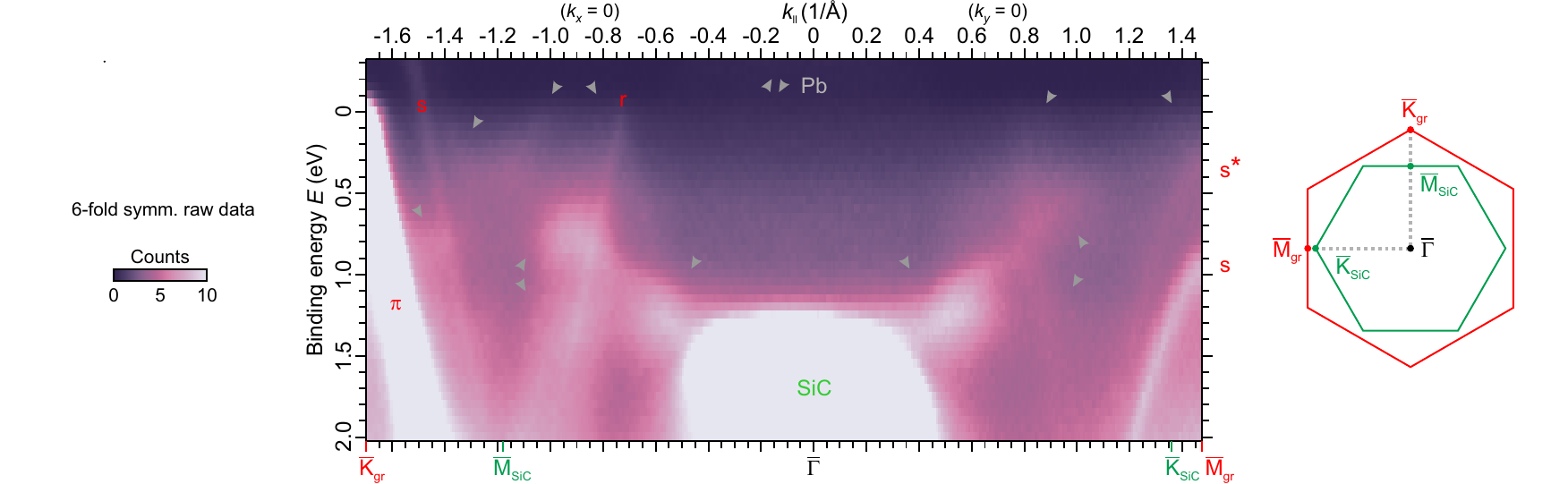}
	\caption{Zoom-in on the low binding energy range of the ARPES energy-momentum cut of Fig.~3(a) in the main manuscript (6-fold symmetrized raw data). The upper threshold of the color scale has also been reduced by a factor of $2$ to further enhance weaker spectral features. Grey triangles highlight Pb-related states.}
	\label{figS2}
\end{figure*}
\vspace{\fill}
\end{document}